\documentclass[prd,aps,a4paper,nofootinbib,twocolumn]{revtex4}  %-1,

\usepackage{graphicx} 
\usepackage{mathrsfs}
\usepackage{amsmath,amsfonts,amssymb}
\usepackage{multirow}

\newcommand{\beq}{\begin{equation}}
\newcommand{\eeq}{\end{equation}}
\begin{document}

\title{Gravitational self-force corrections to tidal invariants for particles on eccentric orbits in a Schwarzschild spacetime}

\author{Donato Bini$^1$ and Andrea Geralico$^1$}
  \affiliation{
$^1$Istituto per le Applicazioni del Calcolo ``M. Picone,'' CNR, I-00185 Rome, Italy
}
 
\date{\today}

\begin{abstract} 
We study tidal effects induced by a particle moving along a slightly eccentric equatorial orbits in a Schwarzschild spacetime within the gravitational self-force framework.
We compute the first order (conservative) corrections in the mass-ratio to the eigenvalues of the electric-type and magnetic-type tidal tensors up to the second order in eccentricity and through the 9.5 post-Newtonian order.
Previous results on circular orbits are thus generalized and recovered in a proper limit.
\end{abstract}

\maketitle

\section{Introduction}

The problem of determining the gravitational field  dynamically  generated by a two-body system in a fully general relativistic context is of great interest today, in view of the recent detections of gravitational wave signals associated with the interactions of (massive, spinning) compact objects \cite{Abbott:2016blz,Abbott:2016nmj,Abbott:2017oio,TheLIGOScientific:2017qsa,Abbott:2017vtc,Abbott:2017gyy}.

However, it is  a matter of fact that 1) no exact solutions of the Einstein's field equations are available yet for an interacting two-body system; 2) numerical relativity (NR) simulations (allowing to enter the strong-field regime of the late-time dynamical evolution of the system) can be performed, but require very long computational times and need to be tested through results obtained by different approaches; 3)
analytical approaches exist only in some limiting situation and use a variety of approximation methods.

During the last years, the Gravitational Self-Force (GSF) approach has proven to be very useful to produce  analytical results valid for extreme mass-ratio binaries and in the weak-field limit of the two-body interaction region (they are indeed achieved in a Post-Newtonian (PN) way, i.e., as series expansion in powers of $1/c$ \cite{Sasaki:2003xr,Barack:2009ux}). 
GSF is, substantially, the most recent development of the general relativistic perturbation theory, containing all necessary tools to explore  the \lq\lq regular" gravitational field generated by one body along the world line of the other one as well as by itself along its own world line.
  
The first explicit GSF analytical calculation was performed in 2013 at the fourth PN level \cite{Bini:2013zaa}. Later on, even more accurate results have been then obtained through (very) high PN-orders, for either metric-based gauge-invariant quantities, like the redshift invariant, connection-based quantities, like the precession angle of a gyroscope, and curvature-based quantities, like the tidal invariants measured along the world line of the perturbing body.
Their  proper transcription  into other formalisms (like the effective one-body (EOB) model introduced by Buonanno and Damour in 1999 \cite{Buonanno:1998gg,Buonanno:2000ef}) has provided a lot of information about the underlying physics during the evolution of the two bodies.

GSF studies were initially performed in the case of circular orbits (corresponding to the initial, inspiralling phase of the interaction) and for structureless particles orbiting a non-spinning (Schwarzschild) black hole \cite{Barack:2010ny,Barack:2009ey,Isoyama:2014mja,Dolan:2013roa,Bini:2014ica,Bini:2015mza,Dolan:2014pja,Bini:2014zxa,Nolan:2015vpa}.  
Generalizations to spinning (Kerr) black holes and non-circular orbits have then been recently obtained in Refs. \cite{Blanchet:2012at,Shah:2012gu,Bini:2015xua,Kavanagh:2016idg,Barack:2011ed,Akcay:2015pza,Bini:2015bfb,Hopper:2015icj,vandeMeent:2015lxa,Tiec:2015cxa,Bini:2016qtx,Bini:2016dvs}.
This kind of calculations are straightforward in principle, but require a lot of technicalities and computational resources making very slow any progress when relaxing the circular condition.

In this paper,  we compute the GSF corrections to the eigenvalues of the electric-type and magnetic-type tidal tensors for slightly eccentric equatorial orbits in the Schwarzschild spacetime up to second order in the eccentricity and through the 9.5PN order. 
Tidal interaction effects have been extensively studied in the literature by using different approaches. The tidal potentials have been identified and (a hierarchy of them) skeletonized through an effective action approach in Ref. \cite{Bini:2012gu}. 
The inclusion of tidal interactions in EOB-based codes computing waveforms for binary neutron star coalescence was done in Refs. \cite{Bernuzzi:2012ci,Bernuzzi:2014owa}, where the comparison between NR long-term simulations and current versions of EOB modeling was discussed, together with  the difficulties in obtaining NR data of sufficient accuracy.
GSF analytical contributions to the tidal dynamics are available only for circular orbits in a Schwarzschild background, in the case of spinless bodies \cite{Bini:2014zxa,Dolan:2014pja,Kavanagh:2015lva,Nolan:2015vpa,Shah:2015nva} and very recently for spinning bodies too, to first order in spin \cite{Bini:2018svh}.
The present work thus provides new information about the two-body tidal interaction in the extreme-mass-ratio limit at disposal of both semi-analytical and fully numerical approaches.
The next step for further research in this context will be 1) the generalization of the present results to a Kerr background, and 2) to consider higher multipolar structure of the perturbing body, in which case available results are limited to spin-square and 1PN approximation \cite{Vines:2010ca,Vines:2011ud,Steinhoff:2016rfi,Vines:2016unv}.

\section{Eccentric orbits and GSF}

Consider a binary system consisting of a spinless compact body (with mass $m_1$) and a Schwarzschild black hole (with mass $m_2$), whose mass ratio is $q \equiv {m_1}/{m_2} \ll 1$.
Through $O(q)$, the small body can then be considered as following a geodesic orbit in a (suitably regularized) perturbed spacetime $g^{\rm R}_{\alpha\beta}$. The latter is decomposed as
\beq
\label{gpert}
g^{\rm R}_{\alpha\beta}=\bar g_{\alpha\beta}+q \, h^{\rm R}_{\alpha\beta} + O(q^2)\,,
\eeq
where $\bar g_{\alpha\beta}$ is the background spacetime 
\begin{eqnarray}
\label{schwmet} 
d\bar s^2 &=&\bar g_{\alpha\beta}dx^\alpha dx^\beta\nonumber\\
&=& -fdt^2+\frac{1}{f}dr^2+r^2(d\theta^2+\sin^2\theta d\phi^2)\,, 
\end{eqnarray}
with $f=1-\frac{2m_2}{r}$, and $q \, h^{\rm R}_{\alpha\beta}$ is the first-order SF metric perturbation. Henceforth, we shall omit the superscript R.

Let the perturbing body move along a slightly eccentric equatorial orbit. The motion is thus characterized by the radial and (averaged) azimuthal angular frequencies
denoted by $\Omega_r=2\pi/T_r$  and $\Omega_{\phi}=\Phi/T_r$, respectively. Here, $T_r$ denotes the radial period, whereas $\Phi$ is the accumulated azimuthal angle from periapsis to periapsis.

Akcay, Dempsey and Dolan \cite{Akcay:2016dku} showed how to calculate the $O(q)$, GSF contribution to 
any gauge-invariant function $Y(m_2\Omega_r, m_2\Omega_\phi; q)$, i.e.,
\begin{eqnarray}
\Delta Y&=& \frac1q \left[Y(m_2\Omega_r,m_2\Omega_{\phi};q)-Y(m_2\Omega_r,m_2\Omega_{\phi};0) \right]\,,\nonumber\\
\end{eqnarray}
for {\it fixed} values of the two frequencies $(\Omega_r,\Omega_{\phi})$. Once the function 
$\Delta Y(\Omega_r,\Omega_{\phi})$ is computed, one can then re-express it as a function of the inverse 
semi latus rectum $u_p=1/p$, and eccentricity $e$, of the unperturbed orbit.

This procedure has been recently applied by Kavanagh et al. \cite{Kavanagh:2017wot} to determine the spin-precession invariant $\Delta \psi(u_p, e)$ up to order $O(e^2)$ in a small-eccentricity expansion, $e\ll1$, and 
up to order $O(u_p^6)$ in the PN expansion, $u_p=1/p\ll1$. 
Their result has been then improved up to the order $O(u_p^9)$ (9PN) in Ref. \cite{Bini:2018aps}.
We compute below the GSF corrections to the eigenvalues of the main tidal tensors (i.e., of the electric and magnetic parts of the Riemann tensor associated with natural observers), generalizing previous results for circular orbits \cite{Bini:2012gu,Bini:2014zxa,Dolan:2014pja,Kavanagh:2015lva,Nolan:2015vpa,Shah:2015nva}.
Our results are accurate to second order in the eccentricity and to the 9.5PN order.
We refer to previous papers \cite{Hopper:2015icj,Kavanagh:2017wot,Bini:2018aps}  for all the necessary details to implement the various steps of the procedure valid also in this case, including the contribution of the gauge, non-radiative modes associated with spherical harmonics characterized by $l=0,1$ and the regularization method. The explicit expression for these mode metrics can be found in Ref. \cite{Hopper:2015icj} (see Appendix A   there).

\subsection{Geodesics in the background spacetime}

The tangent 4-velocity $\bar u$ ($\bar u\cdot\bar u=-1$) to an unperturbed eccentric geodesic orbit on the equatorial plane of the background Schwarzschild spacetime \eqref{schwmet}  with mass $m_2$  is given by 
\beq
\bar u=\bar u^\alpha\partial_\alpha=\frac{\bar E}{f}\partial_t + \dot r \partial_r +\frac{\bar L}{r^2}\partial_\phi\,, 
\eeq
where  $\dot r \equiv \bar u^r$ is such that
\beq
\dot r^2=\left(\frac{dr}{d\bar\tau}\right)^2
=\bar E^2-f\left(1+\frac{\bar L^2}{r^2}\right)\,.
\eeq
Here $\bar E=-\bar u_t$ and $\bar L= \bar u_\phi$ denote the conserved energy and angular momentum per unit mass of the particle, respectively.
The orbit can be parametrized either by the proper time $\bar \tau$ or by the relativistic anomaly $\chi\in[0,2\pi]$, such that 
\beq
r=\frac{m_2 p}{1+e\cos \chi}\,.
\eeq
They are related by 
\beq
\label{dtaudchi}
\frac{d\bar\tau}{d\chi}=\frac{m_2p^{3/2}}{(1+e\cos\chi)^2}\left[\frac{p-3-e^2}{p-6-2e\cos\chi}\right]^{1/2}\,.
\eeq
The (dimensionless) background orbital parameters, semi-latus rectum $p$ and eccentricity $e$, are defined by writing the minimum (pericenter,
$r_{\rm peri}$) and maximum (apocenter, $r_{\rm apo}$) values of the  radial coordinate along the orbit as 
\beq
r_{\rm peri}=\frac{m_2 p}{1+e}\,,\qquad r_{\rm apo}=\frac{m_2 p}{1-e}\,,
\eeq
and are in correspondence with the particle's specific energy and angular momentum via
\beq
\label{energy}
\bar E^2= \frac{ (p-2)^2-4e^2  }{p(p-3-e^2)}\,,\qquad 
\bar L^2=\frac{p^2m_2^2 }{p-3-e^2}\,.
\eeq
The inverse semi latus rectum, $u_p  \equiv p^{-1}$, is  a useful variable, which serves also
as PN expansion parameter.
 
Eq. (\ref{dtaudchi}) can be used to solve the equations for $t$ and $\phi$ as functions of $\chi$, which are 
then expressible in terms of elliptic functions.  
As stated above, eccentric orbits on the background are characterized by two fundamental frequencies,  $\bar\Omega_{r}=2\pi/\bar T_{r}$ and $\bar\Omega_{\phi}=\bar\Phi/\bar T_{r}$, where $\bar\Phi=\oint d\phi = \oint d\chi d\phi/d\chi $ is the angular advance during one radial period, $\bar T_{r}=\oint dt= \oint d\chi dt/d\chi  $. 
To second order in $e$ we find
\begin{eqnarray}
m_2 \bar\Omega_{r} &=& u_p^{3/2}(1-6u_p)^{1/2}\left[1 +\right.\nonumber\\
&-& \left.\frac34 \frac{2-32u_p+165u_p^2-266u_p^3}{ (1-2u_p) (1-6u_p)^2}\, e^2
+O(e^4)\right]\,,\nonumber\\
m_2 \bar\Omega_{\phi} &=& u_p^{3/2}\left[1-\frac32 \frac{1-10u_p+22u_p^2}{(1-2u_p)(1-6u_p)}\, e^2\right.\nonumber\\
&& \left.
+O(e^4)\right]\,.
\end{eqnarray}

\subsection{Geodesics in the perturbed spacetime}

Bound timelike geodesics in the equatorial plane of the perturbed spacetime \eqref{gpert} have $4$-velocity 
\begin{eqnarray}
u&=&u^\alpha\partial_\alpha=(\bar u^\alpha+\delta u^\alpha)\partial_\alpha\\
&=&\frac1{f}(\bar E+\delta E)\partial_t + (\bar u^r+\delta u^r) \partial_r +\frac1{r^2}(\bar L+\delta L)\partial_\phi\,,\nonumber
\end{eqnarray}
with $\delta u^\alpha =O(h)$.
Here, $\delta u^r$ follows from the normalization condition of $u$ with respect to the perturbed metric, which reads
\beq
\label{delta_ur}
\bar u^r \delta u^r = \bar E \delta E -\frac{\bar L}{r^2}f \delta L -\frac12 f h_{00}\,,
\eeq
where $h_{00}=h_{\alpha\beta}\bar u^\alpha \bar u^\beta$.
Equivalently, one can normalize $u$ with respect to the background metric as in Barack and Sago (BS) \cite{Barack:2011ed}, leading to 
\beq
\label{delta_ur_BS}
\bar u^r \delta u^r_{\rm BS} = \bar E \delta E_{\rm BS} -\frac{\bar L}{r^2}f \delta L_{\rm BS}\,,
\eeq
where
\begin{eqnarray}
\label{relwithBS}
\delta E_{\rm BS}&=&\delta E-\frac12\bar E h_{00}\,, \nonumber\\
\delta u^r_{\rm BS}&=&\delta u^r-\frac12\bar u^r h_{00}\,, \nonumber\\
\delta L_{\rm BS}&=&\delta L-\frac12\bar L h_{00}\,.
\end{eqnarray}
The 4-velocity 1-form turns out to be
\begin{eqnarray}
u^\flat&=&u_\alpha dx^\alpha\nonumber\\
&=&-({\bar E}+\delta E-h_{t\bar u})dt +\frac1{f}(\dot r+\delta u^r +fh_{r\bar u}) dr \nonumber\\
&& +({\bar L}+\delta L+h_{\phi \bar u}) d\phi\,,
\end{eqnarray}
where $h_{\alpha\bar u}=h_{\alpha\beta}\bar u^\beta$, and where the further equatorial plane condition $\delta u_\theta=0$ (implying $h_{\theta\bar u}=0$) has been assumed.

The geodesic equations 
\beq
\frac{du_\alpha}{d\tau}-\frac12 (\bar g_{\lambda\mu,\alpha}+h_{\lambda\mu,\alpha})u^\lambda u^\mu=0\,,
\eeq  
determine the evolution of $\delta u_t$ and $\delta u_\phi$, or equivalently of the perturbations in
energy $\delta E$ and angular momentum $\delta L$ by
\begin{eqnarray}
\label{eqdeltaEL}
\frac{d}{d\tau}\delta E &=&\frac12\bar E\frac{dh_{00}}{d\tau}-F_t\,,\nonumber\\  
\frac{d}{d\tau}\delta L &=&\frac12\bar L\frac{dh_{00}}{d\tau}+F_\phi\,,
\end{eqnarray}
where the functions $F_t$ and $F_\phi$ are the covariant $t$ and $\phi$ components of the self force
\beq
F^\mu=-\frac12(\bar g^{\mu\nu}+\bar u^\mu\bar u^\nu)\bar u^\lambda\bar u^\rho(2h_{\nu\lambda;\rho}-h_{\lambda\rho;\nu})\,.
\eeq
Here we are interested in conservative effects only, i.e., we assume that $F^\alpha=F^\alpha_{\rm cons}$ results in a periodic function of $\chi$.
Eqs. \eqref{eqdeltaEL} can then be formally integrated as
\begin{eqnarray}
\delta E_{\rm BS}(\chi) &=& -\int_0^\chi F_t^{\rm cons}(\chi) \frac{d\tau}{d\chi}d\chi+\delta E_{\rm BS}(0)
\nonumber\\
&\equiv& {\mathcal E}_{\rm BS}(\chi)+\delta E_{\rm BS}(0)
\,,\nonumber\\
\delta L_{\rm BS}(\chi) &=& \int_0^\chi F_\phi^{\rm cons}(\chi) \frac{d\tau}{d\chi}d\chi+\delta L_{\rm BS}(0)
\nonumber\\
&\equiv& {\mathcal L}_{\rm BS}(\chi)+\delta L_{\rm BS}(0)
\,,
\end{eqnarray}
recalling the relations \eqref{relwithBS}. Here,  the conservative self force components are defined by $F_t^{\rm cons}=[F_t(\chi)-F_t(-\chi)]/2$ and $F_\phi^{\rm cons}=[F_\phi(\chi)-F_\phi(-\chi)]/2$.
The integration constants $\delta E_{\rm BS}(0)$ and $\delta L_{\rm BS}(0)$ are computed as indicated in Ref. \cite{Barack:2011ed}, and turn out to be
\begin{eqnarray}
\label{deltaELperiBS}
\delta E_{\rm BS}(0)&=& \frac{(1+e)^2 (p-2-2e)}{4e(p-3-e^2)}C
\,,\nonumber\\
\delta L_{\rm BS}(0)&=&  \frac{1}{4e(p-3-e^2)}\frac{C}{B}
\,,
\end{eqnarray}
with
\begin{eqnarray}
B&=&\frac{1}{m_2^2p^3}\, \frac{\bar L}{\bar E}=\frac{1}{m_2p^{3/2}[(p-2)^2-4e^2]^{1/2}}\,,\nonumber\\
C&=&[(1-e)^2(p-2+2e)B {\mathcal L}_{\rm BS}(\pi) -{\mathcal E}_{\rm BS}(\pi)]\,.
\end{eqnarray}

\section{GSF corrections to tidal invariants}

In the perturbed Schwarzschild spacetime, Eq. \eqref{gpert}, the electric-tidal forces are governed by the potential
\begin{eqnarray}
{\rm Tr}[{\mathcal E}(u)^2]&=&{\mathcal E}(u)_{\alpha\beta}{\mathcal E}(u)^{\alpha\beta}\,,\nonumber\\
{\rm Tr}[{\mathcal E}(u)^3]&=&{\mathcal E}(u)_{\alpha\beta}{\mathcal E}(u)^{\beta\mu}{\mathcal E}(u)_{\mu}{}^{\alpha}\,,
\end{eqnarray}
where
\beq
{\mathcal E}(u)_{\alpha\beta}=R_{\alpha\mu\beta\nu}u^\mu u^\nu\,.
\eeq
Similarly, the magnetic-tidal forces follow the potential
\beq
{\rm Tr}[{\mathcal B}(u)^2]={\mathcal B}(u)_{\alpha\beta}{\mathcal B}(u)^{\alpha\beta}\,,
\eeq 
where
\beq
{\mathcal B}(u)_{\alpha\beta}=R^*{}_{\alpha\mu\beta\nu}u^\mu u^\nu\,.
\eeq

We compute below the first-order self-force (1SF) contribution to the eigenvalues of the tidal-electric, and tidal-magnetic, quadrupolar tensors $m_2^2{\mathcal E}(u)^\mu{}_\nu$,  $m_2^2{\mathcal B}(u)^\mu{}_\nu$. These eigenvalues are such that 
\begin{eqnarray}
\label{eq:4.1}
m_2^2 {\mathcal E}(u)&=& {\rm diag} [\lambda_1^{\rm (E)},\lambda_2^{\rm (E)},-(\lambda_1^{\rm (E)}+\lambda_2^{\rm (E)})]\nonumber\\
m_2^2 {\mathcal B}(u)&=& {\rm diag} [\lambda^{\rm (B)},-\lambda^{\rm (B)},0]\,,
\end{eqnarray}
where we used their traceless properties, and the existence of a zero eigenvalue of ${\mathcal B}(u)$ \cite{Dolan:2014pja}. 

Following the prescriptions of Ref. \cite{Akcay:2016dku} we define in the perturbed spacetime the tidal eigenvalues
\beq
\label{lambda_i}
\lambda_i=\frac{\Lambda_i}{{\mathcal T}_r}\,, \qquad
\Lambda_i=\oint\dot\Lambda_i d\tau\,,
\eeq
where ${\mathcal T}_r=\oint d\tau$ denotes the radial period with respect to the proper time $\tau$.
The first-order GSF correction to $\lambda_i$ is then given by
\beq
\Delta \lambda_i=\frac{1}{\bar{\mathcal T}_r}(\Delta \Lambda_i-\bar\Lambda_i\Delta {\mathcal T}_r)
\equiv \lambda^{\rm 1SF}_{i}\,,
\eeq
where
\beq
\Delta \Lambda_i=\delta \Lambda_i-\frac{\partial \bar \Lambda_i}{\partial \bar \Omega_r}\delta \Omega_r -\frac{\partial \bar \Lambda_i}{\partial \bar \Omega_\phi}\delta \Omega_\phi\,,
\eeq
with
\beq
\delta \Lambda_i=\int_0^{2\pi}  \left(\frac{\delta \dot\Lambda_i}{\bar{\dot\Lambda}_i}-\frac{\delta u^r}{\bar u^r}\right) \bar{\dot\Lambda}_i \frac{d \bar \tau}{d\chi}  d\chi\,,
\eeq
where $\bar{\dot\Lambda}_i$ denotes the background value of $d \Lambda_i/d\tau$.
Here $\delta \Omega_r$ and $\delta \Omega_\phi$ denote the GSF corrections to the frequencies 
\beq
\delta \Omega_r=-\bar \Omega_r\frac{\delta T_r}{\bar T_r}\,,\qquad
\delta \Omega_\phi=-\bar \Omega_\phi\left(-\frac{\delta \Phi}{\bar \Phi}+\frac{\delta T_r}{\bar T_r}\right)\,,
\eeq 
where (see Ref. \cite{Akcay:2016dku})
\begin{eqnarray}
\label{deltaTePhi}
\delta T_r&=&\int_0^{2\pi}  \left(\frac{\delta u^t}{\bar u^t}-\frac{\delta u^r}{\bar u^r}\right) \bar u^t \frac{d \bar \tau}{d\chi}  d\chi\nonumber\\
&=&\int_0^{2\pi}  \left(\frac{\delta E}{\bar E}-\frac{\delta u^r}{\bar u^r}\right) \frac{\bar E}{f} \frac{d \bar \tau}{d\chi}  d\chi\,, \nonumber\\
\delta \Phi&=&\int_0^{2\pi}  \left(\frac{\delta u^\phi}{\bar u^\phi}-\frac{\delta u^r}{\bar u^r}\right) \bar u^\phi \frac{d \bar \tau}{d\chi}  d\chi\nonumber\\
&=&\int_0^{2\pi}  \left(\frac{\delta L}{\bar L}-\frac{\delta u^r}{\bar u^r}\right) \frac{\bar L}{r^2} \frac{d \bar \tau}{d\chi}  d\chi\,.
\end{eqnarray}
Finally,
 $\Delta {\mathcal T}_r$ is related to the correction to the Detweiler-Barack-Sago (inverse) redshift function $U=T_r/{\mathcal T}_r$ by
\begin{eqnarray}
\Delta {\mathcal T}_r&=&-\frac{\bar U}{\bar {\mathcal T}_r}\Delta{\mathcal T}_r
=-\frac12\int_0^{\bar {\mathcal T}_r} h_{00}  d\bar\tau \nonumber\\
&=&-\frac12\int_0^{2\pi}h_{00} \frac{d \bar \tau}{d\chi}  d\chi\,.
\end{eqnarray}

We find
\begin{eqnarray}
\lambda^{\rm (E)}_1&=& \lambda^{\rm (E)\, 0SF}_{1 } +q \lambda^{\rm (E)\, 1SF}_{1}\,,\nonumber\\
\lambda^{\rm (E)}_2&=& \lambda^{\rm (E)\,  0SF}_{2} +q \lambda^{\rm (E)\,  1SF}_{2}\,,\nonumber\\
\lambda^{\rm (B)}&=& \lambda^{\rm (B)\, 0SF} +q \lambda^{\rm (B)\, 1SF}\,.
\end{eqnarray}
The unperturbed (0SF) values of these eigenvalues are easily evaluated from Eq. \eqref{lambda_i} with
\begin{eqnarray}
\bar{\dot\Lambda}^{\rm (E)}_1 &=&
-\frac{m_2}{r^5}(2r^2+3\bar L^2)
\,,\nonumber\\
\bar{\dot\Lambda}^{\rm (E)}_2 &=&
\frac{m_2}{r^5}(r^2+3\bar L^2)
\,,\nonumber\\
\bar{\dot\Lambda}^{\rm (B)}&=&\frac{3m_2}{r^5}\bar L(r^2+\bar L^2)^{1/2}
\,,
\end{eqnarray}
and turn out to be given by
\begin{eqnarray}
\lambda^{\rm (E)\, 0SF}_{1}&=&  -u_p^3 \frac{2-3 u_p}{1-3u_p}\nonumber\\
&-&\frac{3u_p^3(99u_p^3-95u_p^2+26u_p-2)}{2(1-3u_p)^2(1-6u_p)}\,e^2\nonumber\\
&+& O(e^4)
\,,\nonumber
\end{eqnarray}
\begin{eqnarray}
\lambda^{\rm (E)\, 0SF}_{2}&=&  \frac{u_p^3}{1-3u_p}\nonumber\\
&+&\frac{3u_p^3(36u_p^3-44u_p^2+13u_p-1)}{2(1-3u_p)^2(1-6u_p)}\,e^2\nonumber\\
&+& O(e^4)
\,,\nonumber\\
-\lambda^{\rm (B)\, 0SF}&=& -3 u_p^{7/2}\frac{\sqrt{1-2 u_p}}{1-3u_p}
+\frac34u_p^{7/2}\times\nonumber\\
&&\frac{594u_p^4-779u_p^3+360u_p^2-67u_p+4}{(1-3u_p)^2(1-6u_p)(1-2u_p)^{3/2}}\,e^2\nonumber\\
&+& O(e^4)
\,.
\end{eqnarray}

The 1SF corrections are
\begin{eqnarray}
\lambda^{\rm (E)\, 1SF}_{1}&=&  \lambda^{\rm (E)\, 1SF}_{1 \, e^0} + e^2  \lambda^{\rm (E)\, 1SF }_{1\, e^2}\,,\nonumber\\
\lambda^{\rm (E)\, 1SF}_{2}&=&  \lambda^{\rm (E)\, 1SF }_{2\, e^0} + e^2  \lambda^{\rm (E)\, 1SF}_{2 \, e^2}\,,\nonumber\\
\lambda^{\rm (B)\,1SF} &=&  \lambda^{\rm (B)\, 1SF }_{e^0} + e^2  \lambda^{\rm (B) \, 1SF }_{ e^2}\,.
\end{eqnarray}

We give below the 1SF corrections to the tidal invariants $\lambda^{\rm (E)\, 1SF}_{1}$, $\lambda^{\rm (E)\, 1SF}_{2}$ and $\lambda^{\rm (B)\,1SF}$ following Refs. \cite{Kavanagh:2017wot,Bini:2018aps}.
Our results are accurate to second order in eccentricity $O(e^2)$ and to 9.5PN order. We postpone to the Appendix the complete, explicit results, anticipating only the first terms of the expansions.

The structure of the various (averaged, as described above) eigenvalues $\lambda^{\rm (E)\, 1SF}_{1}$, $\lambda^{\rm (E)\, 1SF}_{2}$ and $\lambda^{\rm (B)\, 1SF}$ is consistent with results already obtained for the redshift function and the gyroscope precession angle along eccentric orbits, in the sense that we find series of logarithms with polynomial coefficients and an associated (richer and richer) transcendental structure, e.g.,
\begin{eqnarray}
\lambda^{\rm (E)\, 1SF}_{1}&=&P_0(u_p,e^2)+P_1(u_p,e^2)\ln u_p \nonumber\\
&+& P_2(u_p,e^2)\ln^2 u_p+\ldots\,,
\end{eqnarray}
where the functions $P_k(u_p,e^2)$, $k=0,1,2$, have a power expansion in $u_p$ which also involves  fractional powers; the expansion in $e$, instead,  here only involves the two terms $e^0$ and $e^2$; the coefficients of the expansion include terms in $\pi^k$, $\ln^k 2$, $\ln^k 3$, the Euler constant $\gamma^k$, $k=1,2,\ldots $, and products of them  etc. 

We find
\begin{widetext}
\begin{eqnarray}
\lambda^{\rm (E)\, 1SF}_{1 \, e^0}&=&
4 u_p^3+\frac12  u_p^4+\left(-\frac{765}{8}+\frac{123}{32}\pi^2\right) u_p^5\nonumber\\
&&+\left(\frac{635591}{1440}-\frac{1458}{5}\ln(3)-\frac{2512}{15}\gamma-\frac{32909}{3072}\pi^2-\frac{592}{15}\ln(2)-\frac{1256}{15}\ln(u_p)\right) u_p^6
+O_{\ln{}}(u_p^7)\,,
\nonumber\\
\lambda^{\rm (E)\, 1SF}_{1 \, e^2}&=&
-6 u_p^3+\frac{49}{4} u_p^4+\left(\frac{1751}{16}-\frac{615}{128}\pi^2\right) u_p^5\nonumber\\
&&+\left(\frac{75037}{320}+\frac{22599}{5}\ln(3)+\frac{184}{5}\gamma+\frac{28835}{1024}\pi^2-\frac{116312}{15}\ln(2)+\frac{92}{5}\ln(u_p)\right) u_p^6
+O_{\ln{}}(u_p^7)
\,,
\end{eqnarray}

\begin{eqnarray}
\lambda^{\rm (E)\, 1SF}_{2 \, e^0}&=&
-2u_p^3-\frac{19}{4}u_p^4+\left(\frac{625}{16}-\frac{123}{64}\pi^2\right)u_p^5\nonumber\\
&&+\left(-\frac{156251}{2880}-\frac{13945}{6144}\pi^2+\frac{729}{5}\ln(3)+\frac{1256}{15}\gamma+\frac{296}{15}\ln(2)+\frac{628}{15}\ln(u_p)\right)u_p^6
+O_{\ln{}}(u_p^7)\,,
\nonumber\\
\lambda^{\rm (E)\, 1SF}_{2 \, e^2}&=&
3 u_p^3-\frac{19}{8} u_p^4+\left(-\frac{2097}{32}+\frac{615}{256}\pi^2\right) u_p^5\nonumber\\
&&+\left(-\frac{170287}{640}-\frac{8785}{1024}\pi^2-\frac{22599}{10}\ln(3)-\frac{92}{5}\gamma+\frac{58156}{15}\ln(2)-\frac{46}{5}\ln(u_p)\right) u_p^6
+O_{\ln{}}(u_p^7)\,,
\end{eqnarray}
and
\begin{eqnarray}
-\lambda^{\rm (B)\, 1SF}_{e^0}&=&
-7u_p^{7/2}-\frac{7}{4}u_p^{9/2}+\left(-\frac{615}{64}\pi^2+\frac{3509}{16}\right)u_p^{11/2}\nonumber\\
&&
+\left(-\frac{513785}{576}+\frac{628}{3}\ln(u_p)+\frac{115189}{6144}\pi^2+729\ln(3)+\frac{1256}{3}\gamma+\frac{296}{3}\ln(2)\right)u_p^{13/2}
+O_{\ln{}}(u_p^{15/2})\,,
\nonumber\\
-\lambda^{\rm (B)\, 1SF}_{e^2}&=&
\frac{39}{2}u_p^{7/2}+\frac{131}{8}u_p^{9/2}+\left(-\frac{8455}{32}+\frac{2583}{256}\pi^2\right)u_p^{11/2}\nonumber\\
&&
+\left(-\frac{7937887}{5760}-\frac{62}{15}\ln(u_p)-\frac{170351}{3072}\pi^2-\frac{111537}{10}\ln(3)-\frac{124}{15}\gamma+\frac{291076}{15}\ln(2)\right)u_p^{13/2}\nonumber\\
&& +O_{\ln{}}(u_p^{15/2})\,.
\end{eqnarray}

\end{widetext}

The circular orbit limit of these expressions is known with very high accuracy \cite{Kavanagh:2015lva} and is discussed in the next subsection. Concerning the $O(e^2)$ part, the (theoretical) error estimate of our analytical results can be obtained following the analysis presented in Section IV of Ref. \cite{Bini:2018aps}. The eccentricity corrections to $\Delta \lambda_i$ are expected to have the same form as in Eq. (4.5) there, namely
\beq \label{sigmathprev}
\sigma_{N \, \rm PN}^{\rm th}(\Delta \lambda_{i\,e^2}(u_p))= \left|C_{N+\frac12}^{{\Delta \lambda_{i\,e^2}}\!, \,\rm LSO}\right| \frac{(6 u_p)^{(N+\frac12)}}{(1-6u_p)^{\alpha_{N}}} \,,
\eeq
where 
\begin{eqnarray} 
\label{Cnpsi2}
C_{n}^{{\Delta \lambda_{i\,e^2}}\!, \,\rm LSO}
&\equiv& \left[C_n^{{\Delta \lambda_{i\,e^2}}}\!(\ln u_p) u_p^n\right]_{u_p=\frac16} \nonumber\\
&\sim& \pm c({\Delta \lambda_{i\,e^2}})\,,\label{theoryestimate}
\end{eqnarray} 
with $c({\Delta \lambda_{i\,e^2}})$ a number of order unity.
In fact, the presence of the boundary $p=6+2e$ between stable and plunging orbits is likely to introduce a singularity in generic dynamical functions of $p$ and $e$, so that --when expanding such functions in powers of $e$-- a singularity develops at the Last Stable (circular) Orbit (LSO) $u_p=u_{\rm LSO}=1/6$ in the coefficients.
The exponent $\alpha_{N}$ can be suitably chosen to increase the agreement with existing numerical SF data.
We fix it to be zero in absence of available data.
In particular, for $u_p=0.1$ and our 9.5PN expressions this gives roughly $\sigma_{N \, \rm PN}^{\rm th}(\Delta \lambda_{i\,e^2}(u_p))\approx10^{-4}$.
A list of numerical values of the tidal eigenvalues for $e=0.01$ and selected values of $u_p$ is shown in Table \ref{tab:1}, with a number of digits according to the estimate given above.

% table 1

\begin{table}[h]
\centering
\caption{A list of numerical values of the tidal eigenvalues for $e=0.01$ and selected values of $u_p$.
}
\begin{ruledtabular}
\begin{tabular}{cccc}
$u_p$ &  $\lambda^{\rm (E)\, 1SF}_1$ & $\lambda^{\rm (E)\, 1SF}_2$ & $\lambda^{\rm (B)\, 1SF}$  \cr
\hline
0.010 & 0.3999$\, \times 10^{-5}$& -0.2046$\, \times 10^{-5}$& -0.7005$\, \times 10^{-6}$\cr
0.011 &  0.5484$\, \times 10^{-5}$& -0.2812$\, \times 10^{-5}$& -0.1013$\, \times 10^{-5}$\cr
0.013 & 0.7807$\, \times 10^{-5}$& -0.4016$\, \times 10^{-5}$& -0.1529$\, \times 10^{-5}$\cr
0.014 & 0.1165$\, \times 10^{-4}$& -0.6016$\, \times 10^{-5}$& -0.2438$\, \times 10^{-5}$\cr
0.017 &  0.1850$\, \times 10^{-4}$& -0.9600$\, \times 10^{-5}$& -0.4180$\, \times 10^{-5}$\cr
0.020 &  0.3191$\, \times 10^{-4}$& -0.1670$\, \times 10^{-4}$& -0.7910$\, \times 10^{-5}$\cr 
0.025 & 0.6215$\, \times 10^{-4}$& -0.3291$\, \times 10^{-4}$& -0.1721$\, \times 10^{-4}$\cr 
0.033 & 0.1467$\, \times 10^{-3}$& -0.7909$\, \times 10^{-4}$& -0.4686$\, \times 10^{-4}$\cr
0.050 & 0.4867$\, \times 10^{-3}$& -0.2728$\, \times 10^{-3}$& -0.1899$\, \times 10^{-3}$\cr
0.100 & 0.3489$\, \times 10^{-2}$& -0.2177$\, \times 10^{-2}$& -0.1854$\, \times 10^{-2}$\cr
0.111 & 0.4582$\, \times 10^{-2}$& -0.2910$\, \times 10^{-2}$& -0.2520$\, \times 10^{-2}$\cr 
0.125 & 0.6068$\, \times 10^{-2}$& -0.3930$\, \times 10^{-2}$& -0.3409$\, \times 10^{-2}$\cr 
0.143 & 0.7926$\, \times 10^{-2}$& -0.5190$\, \times 10^{-2}$& -0.4403$\, \times 10^{-2}$\cr 
0.167 & 0.9422$\, \times 10^{-2}$& -0.5592$\, \times 10^{-2}$& -0.4329$\, \times 10^{-2}$
\end{tabular}
\end{ruledtabular}
\label{tab:1}
\end{table}

\subsection{Circular orbit limit}

Let us now consider the zero-eccentricity limit of the above expressions.
Akcay et al. \cite{Akcay:2016dku} showed that, in the case of the spin precession invariant, the difference between the limit for vanishing eccentricity of $\Delta\psi$, i.e., lim$_{e\to0}\Delta\psi$, and the corresponding quantity $\Delta\psi^{\rm circ}$ calculated for circular orbits is proportional to the SF correction to the fractional periastron advance $\delta k^{\rm circ}$ (see also Ref. \cite{Akcay:2017azq}), i.e.,
\beq
\lim_{e\to0}\Delta\psi-\Delta\psi^{\rm circ}=\bar G_\psi \delta k^{\rm circ}\,,
\eeq
where
\begin{eqnarray}
\bar G_\psi&=&-\frac{2\pi}{\bar g_1}\frac{\partial \bar \psi}{\partial \bar\Omega_r}\nonumber\\
&=& -\frac{2(1-6u_p)^{5/2}(1-3u_p)^{1/2}}{86u_p^2-39u_p+4}\,,
\end{eqnarray}
with
\beq
\bar g_1=-\frac1{2\pi}\bar T_r\bar\Phi\vert_{e\to0}
= -\frac{2\pi}{u_p^{3/2}(1-6u_p)}\,.
\eeq
Note that the factor $86u_p^2-39u_p+4$ can be traced back to the determinant of the Jacobian matrix $J=\partial(\bar\Omega_r,\bar\Omega_\phi)/\partial(u_p,e)$ in the limit $e\to0$. Indeed
\beq
{\rm det}(J^{-1})\vert_{e\to0}=\frac{4(1-6u_p)^{3/2}(1-2u_p)}{9u_p^3(86u_p^2-39u_p+4)}\,.
\eeq
We recall that $\delta k^{\rm circ}$ is fully known up to the 9.5PN order in terms of the EOB function $\rho$ \cite{Damour:2009sm,Barack:2010ny,Bini:2016qtx}, and is re-expressed here in terms of $u_p$.
Similarly, we find
\begin{eqnarray}
\lim_{e\to0}\Delta\lambda_i-\Delta\lambda_i^{\rm circ}&=&\bar G_{\lambda_i} \delta k^{\rm circ}\,, \nonumber\\
\bar G_{\lambda_i}&=&-\frac{2\pi}{\bar g_1}\frac{\partial \bar \lambda_i}{\partial \bar\Omega_r}\,,
\end{eqnarray}
with 
\begin{eqnarray}
\bar G_{\lambda^{\rm (E)}_1}&=&
\frac{2u_p^2(1-6u_p)^{5/2}(11u_p^2-8u_p+2)}{(86u_p^2-39u_p+4)(1-3u_p)} 
\,,\nonumber\\
\bar G_{\lambda^{\rm (E)}_2}&=&
\frac{2u_p^2(1-6u_p)^{5/2}(-1+4u_p^2)}{(86u_p^2-39u_p+4)(1-3u_p)}  
\,,\nonumber\\
\bar G_{\lambda^{\rm (B)}}&=&
-\frac{5u_p^{5/2}(1-6u_p)^{5/2}(2-9u_p+11u_p^2)}{(86u_p^2-39u_p+4)(1-3u_p)(1-2u_p)^{1/2}}  
\,.\nonumber\\
\end{eqnarray}

\section{Discussion and Outlook}

In the last years several analytical, first-order GSF results along eccentric orbits around a Schwarzschild black hole  have been obtained, concerning redshift, spin precession and gravitational wave energy fluxes \cite{Bini:2015bfb,Hopper:2015icj,Forseth:2015oua,Bini:2016qtx,Kavanagh:2017wot,Bini:2018aps}.
We have computed here 1SF corrections to the eigenvalues of both the electric-type and magnetic-type tidal tensors up to second order in eccentricity $O(e^2)$ and through the 9.5PN order. These results will be of immediate use once converted into other formalisms, like the EOB model, where they will appear as $p_r$-modifications (i.e., eccentricity modifications) to the tidal part of the Hamiltonian (such a tidal part is currently available only in the circular orbit case).
We will face with this transcription in future works. 

Furthermore, possible generalizations of the present analysis --which seem to be at hand-- would require either the use of the Kerr metric instead of the Schwarzschild one, or the addition of the multipolar structure of the perturbing body (tidal invariants in the Kerr case have not been analytically computed yet along circular orbits). We will address these issues in future works too.

\appendix

\section{List of analytical results}

We give below the complete expressions for the 1SF corrections to the electric-type and magnetic-type tidal eigenvalues $\lambda^{\rm (E)\, 1SF}_1$, $\lambda^{\rm (E)\, 1SF}_2$and $\lambda^{\rm (B)\, 1SF}$, to second order in eccentricity $O(e^2)$ and through the 9.5PN order.

\begin{widetext}

\subsection{GSF corrections to $\lambda^{\rm (E)}_1$}

We find
\begin{eqnarray}
\lambda^{\rm (E)\, 1SF}_{1 \, e^0}&=&
4 u_p^3+\frac12  u_p^4+\left(-\frac{765}{8}+\frac{123}{32}\pi^2\right) u_p^5\nonumber\\
&&+\left(\frac{635591}{1440}-\frac{1458}{5}\ln(3)-\frac{2512}{15}\gamma-\frac{32909}{3072}\pi^2-\frac{592}{15}\ln(2)-\frac{1256}{15}\ln(u_p)\right) u_p^6\nonumber\\
&&+\left(\frac{127494449}{28800}-\frac{6837625}{12288}\pi^2-\frac{20876}{105}\ln(2)+\frac{146529}{70}\ln(3)+\frac{137812}{105}\gamma+\frac{68906}{105}\ln(u_p)\right) u_p^7\nonumber\\
&&-\frac{99938}{315}\pi u_p^{15/2}\nonumber\\
&&+\left(\frac{1737491308481}{50803200}+\frac{64769}{945}\gamma-\frac{9765625}{4536}\ln(5)+\frac{3313493}{189}\ln(2)+\frac{7335303}{65536}\pi^4\right.\nonumber\\
&&\left.-\frac{457083}{70}\ln(3)-\frac{6838271315}{1179648}\pi^2+\frac{64769}{1890}\ln(u_p)\right) u_p^8\nonumber\\
&&+\frac{1107097}{450}\pi u_p^{17/2}\nonumber\\
&&+\left(\frac{1354853587056768907}{3520661760000}-\frac{9788863815833}{412876800}\pi^2-\frac{315070534537}{503316480}\pi^4\right.\nonumber\\
&&+\frac{1044921875}{36288}\ln(5)-\frac{1279681043243}{21829500}\gamma+\frac{936036}{175}\ln(3)^2-\frac{2623452402083}{21829500}\ln(2)\nonumber\\
&&-\frac{126976}{15}\zeta(3)-\frac{284734163547}{4312000}\ln(3)+\frac{1862656}{1575}\ln(2)^2+\frac{6793216}{1575}\gamma^2\nonumber\\
&& +\frac{10299392}{1575}\gamma\ln(2)+\frac{1872072}{175}\gamma\ln(3)+\frac{1872072}{175}\ln(2)\ln(3)+\frac{6793216}{1575}\gamma\ln(u_p)\nonumber\\
&&\left.+\frac{936036}{175}\ln(u_p)\ln(3)+\frac{5149696}{1575}\ln(2)\ln(u_p)+\frac{1698304}{1575}\ln(u_p)^2-\frac{1279681043243}{43659000}\ln(u_p)\right) u_p^9\nonumber\\
&&-\frac{3328044031}{1455300}\pi u_p^{19/2}\nonumber\\
&&+\left(-\frac{14489323500013554984923}{2819345937408000}-\frac{678223072849}{46332000}\ln(7)+\frac{235336986894461}{2648646000}\ln(2)\right.\nonumber\\
&&+\frac{985472}{21}\zeta(3)+\frac{1818739765988447}{7945938000}\gamma-\frac{22774784}{735}\gamma^2+\frac{6475857261723}{12556544}\ln(3)\nonumber\\
&&-\frac{5027962488420691}{110981283840}\pi^2-\frac{24562361328125}{145297152}\ln(5)-\frac{56466153}{1225}\ln(3)^2+\frac{201074432}{2205}\ln(2)^2\nonumber\\
&&+\frac{1842207691447831}{32212254720}\pi^4-\frac{112932306}{1225}\ln(2)\ln(3)-\frac{2487808}{2205}\gamma\ln(2)-\frac{112932306}{1225}\gamma\ln(3)\nonumber\\
&&-\frac{56466153}{1225}\ln(u_p)\ln(3)-\frac{1243904}{2205}\ln(2)\ln(u_p)-\frac{22774784}{735}\gamma\ln(u_p)\nonumber\\
&&\left.+\frac{1814061197694047}{15891876000}\ln(u_p)-\frac{5693696}{735}\ln(u_p)^2\right) u_p^{10}\nonumber\\
&&+\left(-\frac{34264742594831201}{349621272000}\pi-\frac{21998344}{4725}\pi^3+\frac{2353822808}{165375}\pi\gamma\right.\nonumber\\
&&\left.+\frac{2002109528}{165375}\pi\ln(2)+\frac{100155852}{6125}\pi\ln(3)+\frac{1176911404}{165375}\pi\ln(u_p)\right) u_p^{21/2}\nonumber
\end{eqnarray}
\begin{eqnarray}
&&+\left(\frac{78517431443756777}{209772763200}\gamma-\frac{3440712074040018511}{8138627481600}\pi^2+\frac{43460925416372417533}{9439774344000}\ln(2)\right.\nonumber\\
&& -\frac{146084607322878663}{69060992000}\ln(3)+\frac{78064993548087017}{419545526400}\ln(u_p)-\frac{2243149779008}{3274425}\gamma\ln(2)\nonumber\\
&& +\frac{4342905234}{13475}\gamma\ln(3)+\frac{4342905234}{13475}\ln(2)\ln(3)+\frac{26182898144}{1091475}\gamma\ln(u_p)\nonumber\\
&& +\frac{2171452617}{13475}\ln(u_p)\ln(3)-\frac{1121574889504}{3274425}\ln(2)\ln(u_p)+\frac{76708984375}{785862}\gamma\ln(5)\nonumber\\
&& +\frac{76708984375}{785862}\ln(2)\ln(5)+\frac{76708984375}{1571724}\ln(u_p)\ln(5)-\frac{1178885630162890625}{3624873348096}\ln(5)\nonumber\\
&& +\frac{3874688}{63}\zeta(3)+\frac{678223072849}{2471040}\ln(7)+\frac{7123234027709651329}{5497558138880}\pi^4\nonumber\\
&& +\frac{2171452617}{13475}\ln(3)^2-\frac{2010562463264}{1403325}\ln(2)^2+\frac{26182898144}{1091475}\gamma^2\nonumber\\
&& +\frac{6545724536}{1091475}\ln(u_p)^2+\frac{76708984375}{1571724}\ln(5)^2-\frac{128148402261}{33554432}\pi^6\nonumber\\
&& \left.-\frac{1746939244497031680332774527}{14513992885776384000}\right) u_p^{11}\nonumber\\
&&+\left(\frac{1028902778}{33075}\pi^3+\frac{270681906956905410287}{520617857760000}\pi-\frac{1300829787}{8575}\pi\ln(3)\right.\nonumber\\
&&\left.-\frac{45071748446}{1157625}\pi\ln(2)-\frac{132397308766}{1157625}\pi\gamma-\frac{66198654383}{1157625}\pi\ln(u_p)\right) u_p^{23/2}\nonumber\\
&&
+O_{\rm ln}(u_p^{12})
\,,  
\end{eqnarray}
and
\begin{eqnarray}
\lambda^{\rm (E)\, 1SF}_{1 \, e^2}&=&
-6 u_p^3+\frac{49}{4} u_p^4+\left(\frac{1751}{16}-\frac{615}{128}\pi^2\right) u_p^5\nonumber\\
&&+\left(\frac{75037}{320}+\frac{22599}{5}\ln(3)+\frac{184}{5}\gamma+\frac{28835}{1024}\pi^2-\frac{116312}{15}\ln(2)+\frac{92}{5}\ln(u_p)\right) u_p^6\nonumber\\
&&+\left(-\frac{93047473}{19200}+\frac{381037}{4096}\pi^2+\frac{617710}{7}\ln(2)-\frac{30200283}{1120}\ln(3)+\frac{12046}{7}\gamma+\frac{6023}{7}\ln(u_p)-\frac{9765625}{672}\ln(5)\right) u_p^7\nonumber\\
&&-\frac{169702}{315}\pi u_p^{15/2}\nonumber\\
&&+\left(\frac{2157688564667}{20321280}+\frac{4843519}{1890}\gamma+\frac{16900390625}{72576}\ln(5)-\frac{1080267617}{1890}\ln(2)\right.\nonumber\\
&&\left.+\frac{58002879}{524288}\pi^4+\frac{139268889}{4480}\ln(3)-\frac{36477234181}{2359296}\pi^2+\frac{4843519}{3780}\ln(u_p)\right) u_p^8\nonumber\\
&&+\frac{3754179661}{352800}\pi u_p^{17/2}\nonumber\\
&&+\left(\frac{289557607057881599}{156473856000}-\frac{12929326735771}{132120576}\pi^2-\frac{40281622425}{67108864}\pi^4\right.\nonumber\\
&&-\frac{20997827734375}{12773376}\ln(5)-\frac{1065943345981}{2910600}\gamma-\frac{17316666}{175}\ln(3)^2-\frac{27855791451491}{43659000}\ln(2)\nonumber\\
&& -41280\zeta(3)+\frac{315105223399929}{137984000}\ln(3)+\frac{480650848}{525}\ln(2)^2+\frac{147232}{7}\gamma^2\nonumber\\
&& +\frac{781076032}{1575}\gamma\ln(2)-\frac{34633332}{175}\gamma\ln(3)-\frac{34633332}{175}\ln(2)\ln(3)+\frac{147232}{7}\gamma\ln(u_p)\nonumber\\
&& -\frac{17316666}{175}\ln(u_p)\ln(3)+\frac{390538016}{1575}\ln(2)\ln(u_p)+\frac{36808}{7}\ln(u_p)^2-\frac{1065943345981}{5821200}\ln(u_p)\nonumber\\
&&\left. -\frac{678223072849}{3041280}\ln(7)\right) u_p^9\nonumber\\
&&-\frac{249893888983}{12700800}\pi u_p^{19/2}\nonumber
\end{eqnarray}
\begin{eqnarray}
&&+\left(-\frac{469668758746869226521617}{28193459374080000}+\frac{54038004663306517}{11860992000}\ln(7)+\frac{1013324799968431}{46332000}\ln(2)\right.\nonumber\\
&& +\frac{16607328}{35}\zeta(3)+\frac{44370615400352249}{15891876000}\gamma-\frac{3291339424}{11025}\gamma^2-\frac{59625253904031261}{3863552000}\ln(3)\nonumber\\
&& +\frac{702054713536051}{8670412800}\pi^2-\frac{13613041834609375}{32546562048}\ln(5)+\frac{613503801}{784}\ln(3)^2-\frac{138395009696}{11025}\ln(2)^2\nonumber\\
&& +\frac{6242120335438697}{42949672960}\pi^4+\frac{613503801}{392}\ln(2)\ln(3)-\frac{302499904}{45}\gamma\ln(2)+\frac{613503801}{392}\gamma\ln(3)\nonumber\\
&& +\frac{613503801}{784}\ln(u_p)\ln(3)-\frac{151249952}{45}\ln(2)\ln(u_p)-\frac{3291339424}{11025}\gamma\ln(u_p)+\frac{44317727237024249}{31783752000}\ln(u_p)\nonumber\\
&& -\frac{822834856}{11025}\ln(u_p)^2+\frac{3173828125}{3528}\ln(2)\ln(5)+\frac{3173828125}{3528}\gamma\ln(5)+\frac{3173828125}{7056}\ln(u_p)\ln(5)\nonumber\\
&& \left.+\frac{3173828125}{7056}\ln(5)^2\right) u_p^{10}\nonumber\\
&&+\left(-\frac{18499284531995386127}{16781821056000}\pi-\frac{42827392}{945}\pi^3+\frac{4582530944}{33075}\pi\gamma+\frac{57319006336}{55125}\pi\ln(2)\right.\nonumber\\
&& \left.-\frac{400623408}{1225}\pi\ln(3)+\frac{2291265472}{33075}\pi\ln(u_p)\right) u_p^{21/2}\nonumber\\
&&+\left(\frac{2424643225400375347}{699242544000}\gamma-\frac{2001617514787851689}{2441588244480}\pi^2-\frac{842192408935393000453}{6293182896000}\ln(2)\right.\nonumber\\
&& +\frac{206357259146535117}{6799851520}\ln(3)+\frac{2412573739696900147}{1398485088000}\ln(u_p)+\frac{45448546210616}{1091475}\gamma\ln(2)\nonumber\\
&& -\frac{760056952041}{431200}\gamma\ln(3)+\frac{132670879857}{61600}\ln(2)\ln(3)+\frac{128474243116}{218295}\gamma\ln(u_p)-\frac{760056952041}{862400}\ln(u_p)\ln(3)\nonumber\\
&& +\frac{22724273105308}{1091475}\ln(2)\ln(u_p)-\frac{194373095703125}{12573792}\gamma\ln(5)-\frac{194373095703125}{12573792}\ln(2)\ln(5)\nonumber\\
&& -\frac{194373095703125}{25147584}\ln(u_p)\ln(5)+\frac{11110161998214161234375}{115995947139072}\ln(5)+\frac{131387296}{315}\zeta(3)\nonumber\\
&& -\frac{6006542031050493541}{142331904000}\ln(7)+\frac{13157240283060929423}{1374389534720}\pi^4-\frac{760056952041}{862400}\ln(3)^2\nonumber\\
&&+\frac{779819447211004}{9823275}\ln(2)^2+\frac{128474243116}{218295}\gamma^2+\frac{32118560779}{218295}\ln(u_p)^2-\frac{194373095703125}{25147584}\ln(5)^2\nonumber\\
&&\left.-\frac{193790772921}{134217728}\pi^6-\frac{138153411052493734048078866847}{145139928857763840000}\right) u_p^{11}\nonumber\\
&&+\left(\frac{21329686963}{37800}\pi^3+\frac{514134786876557740363}{47948060160000}\pi+\frac{1972782394281}{686000}\pi\ln(3)\right.\nonumber\\
&&-\frac{133892971251173}{9261000}\pi\ln(2)-\frac{870570097297}{441000}\pi\gamma-\frac{870570097297}{882000}\pi\ln(u_p)\nonumber\\
&&\left.+\frac{206298828125}{148176}\pi\ln(5)\right) u_p^{23/2}\nonumber\\
&&
+O_{\rm ln}(u_p^{12})
\,.
\end{eqnarray}

\subsection{GSF corrections to $\lambda^{\rm (E)}_2$}

We find
\begin{eqnarray}
\lambda^{\rm (E)\, 1SF}_{2 \, e^0}&=&
-2u_p^3-\frac{19}{4}u_p^4+\left(\frac{625}{16}-\frac{123}{64}\pi^2\right)u_p^5\nonumber\\
&&+\left(-\frac{156251}{2880}-\frac{13945}{6144}\pi^2+\frac{729}{5}\ln(3)+\frac{1256}{15}\gamma+\frac{296}{15}\ln(2)+\frac{628}{15}\ln(u_p)\right)u_p^6\nonumber\\
&&+\left(-\frac{17583369}{6400}+\frac{2316827}{8192}\pi^2-\frac{2070}{7}\gamma+\frac{8034}{35}\ln(2)-\frac{64881}{140}\ln(3)-\frac{1035}{7}\ln(u_p)\right)u_p^7\nonumber\\
&&+\frac{49969}{315}\pi u_p^{15/2}\nonumber\\
&&+\left(-\frac{2642855227157}{101606400}+\frac{9173549735}{2359296}\pi^2-\frac{14822113}{1890}\ln(2)-\frac{3523577}{1890}\gamma\right.\nonumber\\
&&\left.+\frac{9765625}{9072}\ln(5)-\frac{7335303}{131072}\pi^4+\frac{3645}{28}\ln(3)-\frac{3523577}{3780}\ln(u_p)\right) u_p^8
\nonumber\\
&&-\frac{3587807}{6300}\pi u_p^{17/2}\nonumber\\
&&+\left(-\frac{1859700496593276907}{7041323520000}+\frac{2186698067437}{91750400}\pi^2+\frac{1109412659483}{43659000}\ln(2)\right.\nonumber\\
&&+\frac{104811034057}{1006632960}\pi^4+\frac{63488}{15}\zeta(3)+\frac{1086148550243}{43659000}\gamma-\frac{931328}{1575}\ln(2)^2+\frac{343269655947}{8624000}\ln(3)\nonumber\\
&& -\frac{3396608}{1575}\gamma^2-\frac{468018}{175}\ln(3)^2-\frac{244140625}{24192}\ln(5)-\frac{849152}{1575}\ln(u_p)^2+\frac{1086148550243}{87318000}\ln(u_p)\nonumber\\
&& -\frac{936036}{175}\ln(2)\ln(3)-\frac{5149696}{1575}\gamma\ln(2)-\frac{936036}{175}\gamma\ln(3)-\frac{2574848}{1575}\ln(2)\ln(u_p)-\frac{468018}{175}\ln(u_p)\ln(3)\nonumber\\
&&\left. -\frac{3396608}{1575}\gamma\ln(u_p)\right) u_p^9\nonumber\\
&&-\frac{1455120433}{582120}\pi u_p^{19/2}\nonumber\\
&&+\left(\frac{49822242399905561930999}{28193459374080000}-\frac{598221415002293}{21474836480}\pi^4-\frac{359049775283343}{3139136000}\ln(3)\right.\nonumber\\
&& +\frac{92646526117224049}{1109812838400}\pi^2+\frac{678223072849}{92664000}\ln(7)-\frac{214336}{35}\zeta(3)+\frac{1128751953125}{32288256}\ln(5)\nonumber\\
&& +\frac{73404928}{11025}\gamma^2+\frac{6051429}{490}\ln(3)^2-\frac{537966976}{11025}\ln(2)^2-\frac{64177295554583}{15891876000}\gamma\nonumber\\
&& +\frac{79152391926851}{588588000}\ln(2)+\frac{18351232}{11025}\ln(u_p)^2-\frac{59498727260183}{31783752000}\ln(u_p)+\frac{6051429}{245}\gamma\ln(3)\nonumber\\
&& -\frac{9811712}{735}\gamma\ln(2)+\frac{6051429}{245}\ln(2)\ln(3)-\frac{4905856}{735}\ln(2)\ln(u_p)+\frac{6051429}{490}\ln(u_p)\ln(3)\nonumber\\
&& \left.+\frac{73404928}{11025}\gamma\ln(u_p)\right) u_p^{10}\nonumber\\
&&+\left(\frac{96085087293560723}{2097727632000}\pi+\frac{10999172}{4725}\pi^3-\frac{1001054764}{165375}\pi\ln(2)-\frac{50077926}{6125}\pi\ln(3)\right.\nonumber\\
&&\left.-\frac{1176911404}{165375}\pi\gamma-\frac{588455702}{165375}\pi\ln(u_p)\right) u_p^{21/2}\nonumber
\end{eqnarray}
\begin{eqnarray}
&&+\left(-\frac{179851156563883793}{419545526400}\gamma-\frac{39679970206076145349}{18879548688000}\ln(2)-\frac{178893921490849553}{839091052800}\ln(u_p)\right.\nonumber\\
&& +\frac{17963635110262387229}{48831764889600}\pi^2+\frac{38787511868297631}{138121984000}\ln(3)-\frac{186702003}{13475}\ln(2)\ln(3)\nonumber\\
&& +\frac{207218686592}{654885}\gamma\ln(2)-\frac{186702003}{13475}\gamma\ln(3)+\frac{103609343296}{654885}\ln(2)\ln(u_p)-\frac{186702003}{26950}\ln(u_p)\ln(3)\nonumber\\
&& +\frac{7335351712}{218295}\gamma\ln(u_p)-\frac{76708984375}{1571724}\ln(2)\ln(5)-\frac{76708984375}{1571724}\gamma\ln(5)-\frac{76708984375}{3143448}\ln(u_p)\ln(5)\nonumber\\
&& +\frac{225105350542578125}{557672822784}\ln(5)-\frac{28663904}{315}\zeta(3)-\frac{40015161298091}{370656000}\ln(7)\nonumber\\
&& -\frac{8210291604925907841}{10995116277760}\pi^4+\frac{205967661568}{392931}\ln(2)^2+\frac{7335351712}{218295}\gamma^2-\frac{186702003}{26950}\ln(3)^2\nonumber\\
&& +\frac{1833837928}{218295}\ln(u_p)^2-\frac{76708984375}{3143448}\ln(5)^2+\frac{128148402261}{67108864}\pi^6\nonumber\\
&&\left.+\frac{9872329669332193593313996019}{145139928857763840000} \right) u_p^{11}\nonumber\\
&&+\left(-\frac{762062183897159622277}{11453592870720000}\pi-\frac{201872717}{33075}\pi^3-\frac{2159485451}{385875}\pi\ln(2)+\frac{3699785079}{85750}\pi\ln(3)\right.\nonumber\\
&& \left.+\frac{32752736479}{1157625}\pi\gamma+\frac{32752736479}{2315250}\pi\ln(u_p)\right) u_p^{23/2}\nonumber\\
&&
+O_{\rm ln}(u_p^{12})
\,,  
\end{eqnarray}
and
\begin{eqnarray}
\lambda^{\rm (E)\, 1SF}_{2 \, e^2}&=&
3 u_p^3-\frac{19}{8} u_p^4+\left(-\frac{2097}{32}+\frac{615}{256}\pi^2\right) u_p^5\nonumber\\
&&+\left(-\frac{170287}{640}-\frac{8785}{1024}\pi^2-\frac{22599}{10}\ln(3)-\frac{92}{5}\gamma+\frac{58156}{15}\ln(2)-\frac{46}{5}\ln(u_p)\right) u_p^6\nonumber\\
&&+\left(\frac{31747063}{38400}-\frac{179853}{2048}\pi^2-\frac{17697}{35}\gamma-\frac{3010883}{105}\ln(2)+\frac{11992779}{2240}\ln(3)-\frac{17697}{70}\ln(u_p)+\frac{9765625}{1344}\ln(5)\right) u_p^7\nonumber\\
&&+\frac{84851}{315}\pi u_p^{15/2}\nonumber\\
&&+\left(-\frac{8795737412149}{203212800}+\frac{32944511881}{4718592}\pi^2+\frac{107122819}{756}\ln(2)-\frac{20218561}{3780}\gamma-\frac{12681640625}{145152}\ln(5)\right.\nonumber\\
&&\left.-\frac{58002879}{1048576}\pi^4+\frac{36078939}{1792}\ln(3)-\frac{20218561}{7560}\ln(u_p)\right) u_p^8\nonumber\\
&&-\frac{265367029}{78400}\pi u_p^{17/2}\nonumber\\
&&+\left(-\frac{34023816248115041}{34771968000}+\frac{5979953517269}{88080384}\pi^2+\frac{97233078634541}{87318000}\ln(2)+\frac{12899641305}{134217728}\pi^4\right.\nonumber\\
&& +20640\zeta(3)+\frac{306270332257}{1940400}\gamma-\frac{240325424}{525}\ln(2)^2-\frac{312835609037529}{275968000}\ln(3)-\frac{73616}{7}\gamma^2\nonumber\\
&& +\frac{8658333}{175}\ln(3)^2+\frac{172434765625}{405504}\ln(5)-\frac{18404}{7}\ln(u_p)^2+\frac{306270332257}{3880800}\ln(u_p)+\frac{17316666}{175}\ln(2)\ln(3)\nonumber\\
&& -\frac{390538016}{1575}\gamma\ln(2)+\frac{17316666}{175}\gamma\ln(3)-\frac{195269008}{1575}\ln(2)\ln(u_p)+\frac{8658333}{175}\ln(u_p)\ln(3)-\frac{73616}{7}\gamma\ln(u_p)\nonumber\\
&& +\frac{678223072849}{6082560}\ln(7)) u_p^9\nonumber\\
&&-\frac{40688900207}{3628800}\pi u_p^{19/2}\nonumber
\end{eqnarray}
\begin{eqnarray}
&&+\left(\frac{310117288159517228368619}{56386918748160000}-\frac{6309018238623593}{85899345920}\pi^4+\frac{2443492248967143}{702464000}\ln(3)\right.\nonumber\\
&& +\frac{112981482670111187}{1109812838400}\pi^2-\frac{43457724726862117}{23721984000}\ln(7)-\frac{4582128}{35}\zeta(3)\nonumber\\
&&+\frac{166735894103359375}{65093124096}\ln(5)+\frac{1048363184}{11025}\gamma^2-\frac{8234953857}{39200}\ln(3)^2+\frac{1961174704}{441}\ln(2)^2\nonumber\\
&& -\frac{18346664081497487}{31783752000}\gamma-\frac{35803939788195209}{4540536000}\ln(2)+\frac{262090796}{11025}\ln(u_p)^2\nonumber\\
&& -\frac{18293775918169487}{63567504000}\ln(u_p)-\frac{8234953857}{19600}\gamma\ln(3)+\frac{148319648}{63}\gamma\ln(2)-\frac{8234953857}{19600}\ln(2)\ln(3)\nonumber\\
&& +\frac{74159824}{63}\ln(2)\ln(u_p)-\frac{8234953857}{39200}\ln(u_p)\ln(3)+\frac{1048363184}{11025}\gamma\ln(u_p)-\frac{3173828125}{7056}\gamma\ln(5)\nonumber\\
&& \left.-\frac{3173828125}{7056}\ln(2)\ln(5)-\frac{3173828125}{14112}\ln(u_p)\ln(5)-\frac{3173828125}{14112}\ln(5)^2\right) u_p^{10}\nonumber\\
&&+\left(\frac{17893355095703446367}{33563642112000}\pi+\frac{21413696}{945}\pi^3-\frac{28659503168}{55125}\pi\ln(2)\right.\nonumber\\
&&\left.
+\frac{200311704}{1225}\pi\ln(3)-\frac{2291265472}{33075}\pi\gamma-\frac{1145632736}{33075}\pi\ln(u_p)\right) u_p^{21/2}\nonumber\\
&&+\left(-\frac{1567183097539354769}{279697017600}\gamma+\frac{369795179023565387203}{12586365792000}\ln(2)-\frac{1562571900628394129}{559394035200}\ln(u_p)\right.\nonumber\\
&& +\frac{3647820782574901369}{48831764889600}\pi^2+\frac{28567313914065319143}{4419903488000}\ln(3)-\frac{2931955266519}{862400}\ln(2)\ln(3)\nonumber\\
&& -\frac{10173344659516}{1091475}\gamma\ln(2)-\frac{1243202155479}{862400}\gamma\ln(3)-\frac{5086672329758}{1091475}\ln(2)\ln(u_p)\nonumber\\
&&-\frac{1243202155479}{1724800}\ln(u_p)\ln(3)+\frac{261084534818}{1091475}\gamma\ln(u_p)+\frac{3043408203125}{513216}\ln(2)\ln(5)\nonumber\\
&& +\frac{3043408203125}{513216}\gamma\ln(5)+\frac{3043408203125}{1026432}\ln(u_p)\ln(5)-\frac{1386031457101170265625}{33141699182592}\ln(5)\nonumber\\
&& -\frac{301209824}{315}\zeta(3)+\frac{3707061195043663933}{284663808000}\ln(7)-\frac{2703968194942242659}{549755813888}\pi^4\nonumber\\
&& -\frac{35877617299126}{1964655}\ln(2)^2+\frac{261084534818}{1091475}\gamma^2-\frac{1243202155479}{1724800}\ln(3)^2+\frac{130542267409}{2182950}\ln(u_p)^2\nonumber\\
&& \left.+\frac{3043408203125}{1026432}\ln(5)^2+\frac{193790772921}{268435456}\pi^6+\frac{28538684551483258995034063253}{58055971543105536000}\right) u_p^{11}\nonumber\\
&&+\left(-\frac{149924226162309283057}{51636372480000}\pi-\frac{4487269729}{25200}\pi^3+\frac{94800119864213}{18522000}\pi\ln(2)\right.\nonumber\\
&&\left.-\frac{1145495056761}{1372000}\pi\ln(3)+\frac{1769847369859}{2646000}\pi\gamma+\frac{1769847369859}{5292000}\pi\ln(u_p)-\frac{206298828125}{296352}\pi\ln(5)\right) u_p^{23/2}\nonumber\\
&&
+O_{\rm ln}(u_p^{12})
\,.
\end{eqnarray}

\subsection{GSF corrections to $\lambda^{\rm (B)}$}

We find
\begin{eqnarray}
-\lambda^{\rm (B)\, 1SF}_{e^0}&=&
-7u_p^{7/2}-\frac{7}{4}u_p^{9/2}+\left(-\frac{615}{64}\pi^2+\frac{3509}{16}\right)u_p^{11/2}\nonumber\\
&&
+\left(-\frac{513785}{576}+\frac{628}{3}\ln(u_p)+\frac{115189}{6144}\pi^2+729\ln(3)+\frac{1256}{3}\gamma+\frac{296}{3}\ln(2)\right)u_p^{13/2}\nonumber\\
&&
+\left(-\frac{147443}{105}\ln(u_p)-\frac{141647189}{11520}+\frac{35413007}{24576}\pi^2-\frac{136323}{28}\ln(3)-\frac{294886}{105}\gamma+\frac{22562}{21}\ln(2)\right)u_p^{15/2}\nonumber\\
&&
+\frac{49969}{63}\pi u_p^8\nonumber\\
&&
+\left(-\frac{4674463}{3780}\ln(u_p)-\frac{9434749494979}{101606400}-\frac{4674463}{1890}\gamma-\frac{2427613}{54}\ln(2)+\frac{48828125}{9072}\ln(5)\right.\nonumber\\
&&\left.
+\frac{110079}{8}\ln(3)+\frac{36091610815}{2359296}\pi^2-\frac{36676515}{131072}\pi^4\right)u_p^{17/2}\nonumber\\
&&
-\frac{2303671}{420}\pi u_p^9\nonumber\\
&&
+\left(-\frac{936036}{35}\gamma\ln(3)-\frac{5149696}{315}\gamma\ln(2)-\frac{936036}{35}\ln(2)\ln(3)-\frac{1387177929347162137}{1408264704000}\right.\nonumber\\
&&
+\frac{1252442084573}{17463600}\ln(u_p)+\frac{1252442084573}{8731800}\gamma+\frac{2421022877093}{8731800}\ln(2)-\frac{5029296875}{72576}\ln(5)\nonumber\\
&&
+\frac{291423720267}{1724800}\ln(3)+\frac{63488}{3}\zeta(3)+\frac{3560963628011}{55050240}\pi^2+\frac{298205268073}{201326592}\pi^4-\frac{3396608}{315}\gamma^2\nonumber\\
&&
-\frac{931328}{315}\ln(2)^2-\frac{468018}{35}\ln(3)^2-\frac{849152}{315}\ln(u_p)^2-\frac{3396608}{315}\gamma\ln(u_p)-\frac{2574848}{315}\ln(2)\ln(u_p)\nonumber\\
&&\left.
-\frac{468018}{35}\ln(u_p)\ln(3)\right)u_p^{19/2}\nonumber\\
&&
+\frac{317855299}{207900}\pi u_p^{10}\nonumber\\
&&
+\left(\frac{64238456}{3675}\ln(u_p)^2-\frac{2493148144963403}{10594584000}\ln(u_p)+\frac{53190027}{490}\ln(u_p)\ln(3)-\frac{154079168}{11025}\gamma\ln(2)\right.\nonumber\\
&&
+\frac{53190027}{245}\gamma\ln(3)+\frac{53190027}{245}\ln(2)\ln(3)+\frac{356890910670224675213741}{28193459374080000}-\frac{2500945758787403}{5297292000}\gamma\nonumber\\
&&
+\frac{43947124246410797}{369937612800}\pi^2-\frac{618657322046537}{15891876000}\ln(2)+\frac{37543134765625}{96864768}\ln(5)-\frac{2624809696}{11025}\ln(2)^2\nonumber\\
&&
+\frac{53190027}{490}\ln(3)^2-\frac{752030326906821}{627827200}\ln(3)+\frac{678223072849}{18532800}\ln(7)-\frac{1843139939552983}{12884901888}\pi^4\nonumber\\
&&\left.
-\frac{10763392}{105}\zeta(3)+\frac{256953824}{3675}\gamma^2+\frac{256953824}{3675}\gamma\ln(u_p)-\frac{77039584}{11025}\ln(2)\ln(u_p)\right)u_p^{21/2}\nonumber\\
&&
+\left(-\frac{588455702}{33075}\pi\ln(u_p)-\frac{1176911404}{33075}\pi\gamma-\frac{50077926}{1225}\pi\ln(3)-\frac{1001054764}{33075}\pi\ln(2)\right.\nonumber\\
&&\left.
+\frac{103117720570197691}{419545526400}\pi+\frac{10999172}{945}\pi^3\right)u_p^{11}\nonumber
\end{eqnarray}
\begin{eqnarray}
&&
+\left(-\frac{16741141032149288291}{1452272976000}\ln(2)+\frac{90598863811273767}{19731712000}\ln(3)-\frac{2638248299742475703}{2097727632000}\gamma\right.\nonumber\\
&&
-\frac{2624963810194546103}{4195455264000}\ln(u_p)+\frac{46135491239131491559}{48831764889600}\pi^2+\frac{640742011305}{67108864}\pi^6\nonumber\\
&&
-\frac{383544921875}{3143448}\ln(5)^2-\frac{524492901}{770}\gamma\ln(3)+\frac{1143786659392}{654885}\gamma\ln(2)-\frac{524492901}{770}\ln(2)\ln(3)\nonumber\\
&&
-\frac{11764078976}{1091475}\gamma\ln(u_p)+\frac{571893329696}{654885}\ln(2)\ln(u_p)-\frac{524492901}{1540}\ln(u_p)\ln(3)\nonumber\\
&&
-\frac{383544921875}{1571724}\ln(2)\ln(5)-\frac{383544921875}{1571724}\gamma\ln(5)-\frac{383544921875}{3143448}\ln(u_p)\ln(5)\nonumber\\
&&
+\frac{7153039085224609375}{7249746696192}\ln(5)-\frac{69917536}{315}\zeta(3)-\frac{49510284317977}{74131200}\ln(7)-\frac{36565261546718516997}{10995116277760}\pi^4\nonumber\\
&&
-\frac{11764078976}{1091475}\gamma^2+\frac{34338579190688}{9823275}\ln(2)^2-\frac{524492901}{1540}\ln(3)^2-\frac{2941019744}{1091475}\ln(u_p)^2\nonumber\\
&&\left.
+\frac{45037622187610223222383061623}{145139928857763840000}\right)u_p^{23/2}\nonumber\\
&&
+\left(\frac{33923529793}{128625}\pi\gamma+\frac{84987239753}{1157625}\pi\ln(2)+\frac{6153603453}{17150}\pi\ln(3)+\frac{33923529793}{257250}\pi\ln(u_p)\right.\nonumber\\
&&\left.
-\frac{2618939690282810274271}{2290718574144000}\pi-\frac{259138099}{3675}\pi^3\right)u_p^{12}\nonumber\\
&&
+O_{\rm ln}(u_p^{25/2})
\,,  
\end{eqnarray}
and
\begin{eqnarray}
-\lambda^{\rm (B)\, 1SF}_{e^2}&=&
\frac{39}{2}u_p^{7/2}+\frac{131}{8}u_p^{9/2}+\left(-\frac{8455}{32}+\frac{2583}{256}\pi^2\right)u_p^{11/2}\nonumber\\
&&
+\left(-\frac{7937887}{5760}-\frac{62}{15}\ln(u_p)-\frac{170351}{3072}\pi^2-\frac{111537}{10}\ln(3)-\frac{124}{15}\gamma+\frac{291076}{15}\ln(2)\right)u_p^{13/2}\nonumber\\
&&
+\left(-\frac{362221}{210}\ln(u_p)+\frac{362747239}{115200}+\frac{63041}{1536}\pi^2+\frac{29290491}{448}\ln(3)-\frac{362221}{105}\gamma-\frac{22302893}{105}\ln(2)\right.\nonumber\\
&&\left.
+\frac{48828125}{1344}\ln(5)\right)u_p^{15/2}\nonumber\\
&&
+\frac{474224}{315}\pi u_p^8\nonumber\\
&&
+\left(-\frac{6065461}{840}\ln(u_p)-\frac{20148288166799}{67737600}-\frac{6065461}{420}\gamma+\frac{727694981}{540}\ln(2)-\frac{27236328125}{48384}\ln(5)\right.\nonumber\\
&&\left.
-\frac{653877279}{8960}\ln(3)+\frac{64466993717}{1572864}\pi^2-\frac{348696819}{1048576}\pi^4\right)u_p^{17/2}\nonumber\\
&&
-\frac{5577450491}{235200}\pi u_p^9\nonumber\\
&&
+\left(\frac{85647294}{175}\gamma\ln(3)-\frac{1957839776}{1575}\gamma\ln(2)+\frac{85647294}{175}\ln(2)\ln(3)-\frac{67766206056353335589}{14082647040000}\right.\nonumber\\
&&
+\frac{77470591458361}{174636000}\ln(u_p)+\frac{678223072849}{1216512}\ln(7)+\frac{77470591458361}{87318000}\gamma+\frac{158295935863721}{87318000}\ln(2)\nonumber\\
&&
+\frac{33094009765625}{8515584}\ln(5)-\frac{1543957109009541}{275968000}\ln(3)+\frac{1611488}{15}\zeta(3)+\frac{572506304520313}{2202009600}\pi^2\nonumber\\
&&
+\frac{4838989091909}{2013265920}\pi^4-\frac{86214608}{1575}\gamma^2-\frac{3605812688}{1575}\ln(2)^2+\frac{42823647}{175}\ln(3)^2-\frac{21553652}{1575}\ln(u_p)^2\nonumber\\
&&\left.
-\frac{86214608}{1575}\gamma\ln(u_p)-\frac{978919888}{1575}\ln(2)\ln(u_p)+\frac{42823647}{175}\ln(u_p)\ln(3)\right)u_p^{19/2}\nonumber
\end{eqnarray}
\begin{eqnarray}
&&
+\frac{4804820710843}{279417600}\pi u_p^{10}\nonumber\\
&&
+\left(\frac{17902154}{105}\ln(u_p)^2-\frac{2439564222568427}{825552000}\ln(u_p)-\frac{74290603221}{39200}\ln(u_p)\ln(3)+\frac{59326710256}{3675}\gamma\ln(2)\right.\nonumber\\
&&
-\frac{74290603221}{19600}\gamma\ln(3)-\frac{74290603221}{19600}\ln(2)\ln(3)+\frac{13929241577707319798017}{322210964275200}-\frac{2443120040142827}{412776000}\gamma\nonumber\\
&&
-\frac{7236628125540281}{31708938240}\pi^2-\frac{147403168764033517}{2889432000}\ln(2)+\frac{123463883278515625}{65093124096}\ln(5)\nonumber\\
&&
+\frac{47619891416}{1575}\ln(2)^2-\frac{74290603221}{39200}\ln(3)^2+\frac{3690708045527111967}{100452352000}\ln(3)\nonumber\\
&&
-\frac{263403723249605491}{23721984000}\ln(7)-\frac{103596443546418371}{257698037760}\pi^4-\frac{3159008}{3}\zeta(3)+\frac{71608616}{105}\gamma^2\nonumber\\
&&
-\frac{15869140625}{14112}\ln(5)^2-\frac{15869140625}{7056}\gamma\ln(5)-\frac{15869140625}{7056}\ln(2)\ln(5)-\frac{15869140625}{14112}\ln(u_p)\ln(5)\nonumber\\
&&\left.
+\frac{71608616}{105}\gamma\ln(u_p)+\frac{29663355128}{3675}\ln(2)\ln(u_p)\right)u_p^{21/2}\nonumber\\
&&
+\left(-\frac{596515798}{3375}\pi\ln(u_p)+\frac{13331070926528518129}{4794806016000}\pi-\frac{1193031596}{3375}\pi\gamma-\frac{430893602284}{165375}\pi\ln(2)\right.\nonumber\\
&&\left.
+\frac{78048796}{675}\pi^3+\frac{4957714674}{6125}\pi\ln(3)\right)u_p^{11}\nonumber\\
&&
+\left(\frac{11401105010656344200029}{37759097376000}\ln(2)-\frac{295242793024246690167}{4419903488000}\ln(3)-\frac{53926547475927154907}{4195455264000}\gamma\right.\nonumber\\
&&
-\frac{53712733650216870107}{8390910528000}\ln(u_p)+\frac{41061722900238542027}{48831764889600}\pi^2+\frac{1481547473649}{268435456}\pi^6\nonumber\\
&&
+\frac{34902197265625}{1862784}\ln(5)^2+\frac{611863203267}{172480}\gamma\ln(3)-\frac{314174044715108}{3274425}\gamma\ln(2)\nonumber\\
&&
-\frac{1076889907773}{172480}\ln(2)\ln(3)-\frac{11979998546}{14553}\gamma\ln(u_p)-\frac{157087022357554}{3274425}\ln(2)\ln(u_p)\nonumber\\
&&
+\frac{611863203267}{344960}\ln(u_p)\ln(3)+\frac{34902197265625}{931392}\ln(2)\ln(5)+\frac{34902197265625}{931392}\gamma\ln(5)\nonumber\\
&&
+\frac{34902197265625}{1862784}\ln(u_p)\ln(5)+\frac{239879662969024977124531465021}{96759952571842560000}-\frac{1984320321287159921875}{8592292380672}\ln(5)\nonumber\\
&&
-\frac{64620088}{35}\zeta(3)+\frac{2608076223554828089}{25878528000}\ln(7)-\frac{821649941639315317127}{32985348833280}\pi^4-\frac{11979998546}{14553}\gamma^2\nonumber\\
&&\left.
-\frac{54773937953942}{297675}\ln(2)^2+\frac{611863203267}{344960}\ln(3)^2-\frac{5989999273}{29106}\ln(u_p)^2\right)u_p^{23/2}\nonumber\\
&&
+\left(-\frac{4445360816902145649492787}{183257485931520000}\pi-\frac{230652027331}{176400}\pi^3+\frac{85831564052201}{18522000}\pi\gamma+\frac{214546784763811}{6174000}\pi\ln(2)\right.\nonumber\\
&&\left.
-\frac{9550592868573}{1372000}\pi\ln(3)+\frac{85831564052201}{37044000}\pi\ln(u_p)-\frac{1031494140625}{296352}\pi\ln(5)\right)u_p^{12}\nonumber\\
&&
+O_{\rm ln}(u_p^{25/2})
\,.
\end{eqnarray}

\end{widetext}

\section*{Acknowledgments}

The authors thank T. Damour for useful discussions.
D.B. thanks the Naples Section of the Italian Istituto Nazionale di Fisica Nucleare (INFN) and the International Center for Relativistic Astrophysics Network (ICRANet) for partial support.

\end{document}